\documentclass[prd,aps,showpacs,tightenlines,nofootinbib,preprint,preprintnumbers]{revtex4}
\usepackage{graphicx}
\usepackage{amsfonts}
\usepackage{amssymb}
\usepackage{latexsym}
\usepackage{cancel}
\usepackage{color}

\newcommand{\bear}{\begin{array}}  \newcommand{\eear}{\end{array}}
\newcommand{\bea}{\begin{eqnarray}}  \newcommand{\eea}{\end{eqnarray}}
\newcommand{\beq}{\begin{equation}}  \newcommand{\eeq}{\end{equation}}
\newcommand{\bef}{\begin{figure}}  \newcommand{\eef}{\end{figure}}
\newcommand{\bec}{\begin{center}}  \newcommand{\eec}{\end{center}}
\newcommand{\non}{\nonumber}  
\newcommand{\lmk}{\left(}  \newcommand{\rmk}{\right)}

\newcommand{\del}{\partial}  
\newcommand{\vect}[1]{\mbox{\boldmath${#1}$}}
\newcommand{\bib}{\bibitem} 
\newcommand{\la}{\left\langle} \newcommand{\ra}{\right\rangle}

\def\IB#1#2#3{{\bf #1}, #2 (19#3)}
\def\IBB#1#2#3{{\bf #1}, #2 (20#3)}
\def\IBID#1#2#3{{\it ibid}. {\bf #1}, #2 (19#3)}
\def\IBIDD#1#2#3{{\it ibid}. {\bf #1}, #2 (20#3)}

\def\AAA#1#2#3{Astron. Astrophys. {\bf #1}, #2 (20#3)}

\def\APJJ#1#2#3{Astrophys. J. {\bf #1}, #2 (20#3)}

\def\APJSS#1#2#3{Astrophys. J. Suppl. {\bf #1}, L#2 (20#3)}

\def\JCAPP#1#2#3{J. Cosmol. Astropart. Phys. {\bf #1}, #2 (20#3)}

\def\JHEPP#1#2#3{J. High Energy Phys. {\bf #1}, #2 (20#3)}

\def\JP#1#2#3{J. Phys. A {\bf #1}, #2 (19#3)}

\def\MNRASS#1#2#3{Mon. Not. R. Astron. Soc. {\bf #1}, #2 (20#3)}

\def\NPB#1#2#3{Nucl. Phys. {\bf B#1}, #2 (19#3)}

\def\PLB#1#2#3{Phys. Lett. B {\bf #1}, #2 (19#3)}
\def\PLBB#1#2#3{Phys. Lett. B {\bf #1}, #2 (20#3)}
\def\PLBold#1#2#3{Phys. Lett. {\bf#1B}, #2 (19#3)}

\def\PRD#1#2#3{Phys. Rev. D {\bf #1}, #2 (19#3)}
\def\PRDD#1#2#3{Phys. Rev. D {\bf #1}, #2 (20#3)}

\def\PRL#1#2#3{Phys. Rev. Lett. {\bf#1}, #2 (19#3)}

\def\PTP#1#2#3{Prog. Theor. Phys. {\bf #1}, #2 (19#3)}

\def\RPP#1#2#3{Rep. Prog. Phys. {\bf #1}, #2 (19#3)}

\begin{document}

\hfill hep-ph/0606287\\

\title{Implications of cosmic strings with time-varying tension  \\
on CMB and large scale structure}
%
\author{\vspace{0.5cm}
Kazuhide Ichikawa$^{1}$, Tomo Takahashi$^{2}$ and Masahide Yamaguchi$^{3}$
\vspace{0.5cm}}

\affiliation{
$^{1}$Institute for Cosmic Ray Research,
University of Tokyo,
Kashiwa 277-8582, Japan, \\
$^{2}$Department of physics, Saga University, Saga 840-8502, Japan,  \\
$^{3}$Department of Physics and Mathematics, Aoyama Gakuin
University, Sagamihara 229-8558, Japan
\vspace{2cm}
}

%

\begin{abstract}
We investigate cosmological evolution and implications of cosmic strings
with time-dependent tension. We derive basic equations of time
development of the correlation length and the velocity of such strings,
based on the one scale model. Then, we find that, in the case  where the
tension depends on some power of the cosmic time, cosmic strings with
time-dependent tension goes into the scaling solution if the power is
lower than a critical value. We also discuss  cosmic
microwave background anisotropy and matter power spectra produced by
these strings.  The constraints on their tensions from the
Wilkinson microwave anisotropy probe (WMAP) three year data and Sloan
digital sky survey (SDSS) data are also given.
\end{abstract}


\maketitle


\section{Introduction}

Topological defects can be produced as a result of thermal
\cite{Kibble} or nonthermal \cite{KVY} phase transitions at the
symmetry breaking in the early universe. Among them, cosmic strings
have been extensively investigated mainly because they could act as
the seeds for large scale structure. Recent observations of the cosmic
microwave background (CMB) anisotropy \cite{WMAP} reveal that cosmic
strings cannot be the primary source of primordial density
fluctuations, and give the bound on the tension of cosmic strings
$G\mu < 2.7 \times 10^{-7}$ at $95\%$ confidence level \cite{PWW},
which is obtained by considering a hybrid scenario of adiabatic
fluctuations generated during inflation plus isocurvature fluctuations
produced by cosmic strings. The constraint given in other analysis
such as Refs.~\cite{Fraisse} and \cite{Seljak} are also in good
agreement with theirs.

However, cosmic strings can still affect various astrophysics
\cite{VSHK} such as the early reionization \cite{ER}, gravitational
radiation \cite{GR}, gravitational lensing effects \cite{GL}, the
neutrino masses \cite{BMY}, and so on. Furthermore, an interesting
possibility is revived that fundamental strings of the superstring
theory can be expanded to cosmological sizes and act as cosmic strings
\cite{CSS}. In fact, F-strings and D-strings are produced at the end of
brane inflation \cite{BI}.

In almost all research so far, the tensions of cosmic strings have
been assumed to be constant. Recently however, one of the authors
(M.Y.) pointed out that tensions of cosmic strings can depend on the
cosmic time \cite{MY}. For example, a complex scalar field $\phi$,
which allows a string solution, couples to another field $\chi$,
\beq
  V(\phi,\chi) = \frac{\lambda}{4}(|\phi|^2 - \chi^2)^2 
                + \frac12 m_{\chi}^2 \chi^2.
\eeq
When a coupling constant $\lambda$ is small enough, the backreaction to
the oscillation of $\chi$ is negligible. Then, tension of a string $\mu$
is determined by the root mean square of the expectation value of $\chi$
and given by $\mu \propto a^{-3}$ ($a$:the scale factor), which is
proportional to $t^{-3/2}$ in the radiation dominated era and $t^{-2}$
in the matter dominated era. Hence, the tension $\mu$ depends on the
power of the cosmic time. Furthermore, in the warped geometry, the
tension depends on the position of the brane in the bulk, which can
depend on the cosmic time before the radion is fixed completely. Thus,
it is quite natural to consider cosmic strings with time-dependent
tension.

The key property of cosmological evolution of cosmic strings is scaling,
which is confirmed by extensive investigations of cosmic strings with
constant tension \cite{VSHK}. It has been shown that the cosmic string
network goes into the scaling regime, in which the typical length of the
cosmic string network grows with the horizon scale. Then, the number of
infinite strings per horizon volume is a constant irrespective of time
and hence the ratio of the energy density of infinite strings to that of
the background universe is constant. Thus, cosmic strings can generate
scale invariant density fluctuations. Scaling property of the cosmic
string network is confirmed both analytically \cite{Kibble} and
numerically \cite{AT,BB,AS} by using the Nambu-Goto action \cite{NG},
which can be obtained after integrating out heavy modes of particles and
neglecting high curvature of the geometry.\footnote{Scaling property of
the evolution is known not only for cosmic (local) strings but also for
global strings \cite{gs} and global monopoles \cite{gm}.}

As for cosmic strings with time-dependent tension, the followings are
shown in Ref.~\cite{MY}.  First of all, it is shown that the effective
action of a cosmic string with time-dependent tension is given by the
Nambu-Goto action with an additional factor for the time-dependent
tension. By making use of such a Nambu-Goto-like action, the equation of
motion in an expanding universe is derived and the evolution of cosmic
strings with time-dependent tension is investigated. Then, it is
confirmed that, in the case where the tension changes as the power $q$
of time, the string network goes into the scaling regime, in which the
characteristic scale of the string network grows in proportion to the
cosmic time. One should notice that the ratio of the energy density of
infinite strings to that of the background universe is {\it not}
necessarily constant due to the time dependence of the tension,
different from the case of cosmic strings with constant tension.
However, in Ref.~\cite{MY}, the constancy of string velocity is
implicitly assumed to derive the scaling solution. In this paper, we
also derive the equation of time development of string velocity based on
the velocity-dependent one scale model \cite{MS} and find a critical
power of time dependence of the tension, below which the scaling
property is realized and string velocity becomes almost constant.

When the tension of cosmic strings depends on time, its effects on
astrophysics and cosmology can be significantly different from those
of the conventional cosmic strings.  Thus we should consider the
implications of such cosmic strings on various astrophysical and
cosmological issues.  Among them, we study its effects on CMB and
large scale structure in this paper.  Since the time dependence of the
tension strongly depends on models, we investigate it in some general
settings.  Here we assume the time dependence of the tension as $\mu
\propto a^n$ and $\mu \propto \tau^n$ where $\tau$ is the conformal
time and $n$ parametrize the dependence on the power.  With this
assumption, we study the CMB and matter power spectrum in models with
the time dependent tension. For this purpose, we modified the CMBACT
code developed and made publicly available by Pogosian and Vachaspati
\cite{PV,PWW} to calculate the CMB anisotropy and the matter power
spectrum induced by cosmic strings with time dependent tension. We
will also give the constraint on the cosmic strings by comparing the
predictions with recent cosmological observations, especially, the
WMAP three-year data \cite{WMAP} and the SDSS data \cite{SDSS}.  We
would like to stress that these constraints are important in
discussing astrophysical applications such as the early reionization,
gravitational radiation, gravitational lensing effects, the neutrino
masses, and so on.

This paper is organized as follows. In the next section, we derive basic
equations to investigate cosmological evolution of cosmic strings with
time-dependent tension and obtain the condition under which the string
network goes into the scaling solution. In section III, we calculate the
predictions of the CMB anisotropy and the matter power spectrum induced
by cosmic strings with time-dependent tension and give the constraints
on their tensions from the WMAP three year results and SDSS data. Since
the time dependence of the tension strongly depends on a model, we
concentrate on the case that the tension changes in proportion to the
power of the conformal time or the scale factor. The final section is 
devoted to conclusions and discussion.

\section{Cosmological evolution of cosmic strings with 
time-dependent tension}

As shown in Ref. \cite{MY}, the effective action for a cosmic string
with time-dependent tension is given by
\beq
  S_{\rm eff} = - \int d^2 \zeta \sqrt{-\gamma} \,\mu(\tau).
\eeq
Here, two parameters $\zeta^a$($a = 0, 1$) characterize the worldsheet
swept by a cosmic string with $x^{\mu} = x^{\mu}(\zeta^{a})$.  We take
the timelike coordinate $\zeta^0$ to be the cosmic conformal time $\tau$
and the spacelike coordinate $\zeta^1$ to be $\sigma$, which
parametrizes the string at a fixed time. The metric is taken to be that
of the spatially flat expanding universe,
\beq
  ds^2 = g_{\mu\nu} dx^{\mu} dx^{\nu}
       = dt^2 - a^2(t) d\vect x^2
       = a^2(\tau) (d\tau^2 - d\vect x^2)
\eeq
with $d\tau = dt / a(t)$. $\gamma_{ab} \equiv g_{\mu\nu} x^{\mu}_{,a}
x^{\nu}_{,b}$ is the spacetime metric on the string worldsheet,
\bea
  ds^2 &=& g_{\mu\nu} dx^{\mu} dx^{\nu} \non \\
       &=& \gamma_{ab} d\zeta^{a} d\zeta^{b}.
\eea
Here, the comma represents the partial derivative. Since we are
interested only in the transverse motion of cosmic strings, the metric
should satisfy the following condition,
\beq
  \dot{\vect{x}} \cdot \vect{x}' = 0 \quad
  \Longleftrightarrow \quad \gamma_{01}=\gamma_{10}=0,
\eeq
where dots and primes represent derivatives with respect to conformal
time $\tau$ and the spacelike parameter $\sigma$, respectively.

The Euler-Lagrange equation for the effective action is given by
\beq
   \frac{\mu}{\sqrt{-\gamma}} 
      \del_{a}(\sqrt{-\gamma} \gamma^{ab} x^{\mu}_{b})
   + \mu^{,a} x^{\mu}_{,a} 
   + \mu \Gamma^{\mu}_{\nu \sigma} \gamma^{ab} 
     x^{\sigma}_{,a} x^{\nu}_{,b} = 0,
\eeq
where $\Gamma^{\mu}_{\nu \sigma}$ is the four-dimensional Christoffel
symbol given by
\beq
  \Gamma^{\mu}_{\nu \sigma} = \frac12 g^{\mu \tau}
    (g_{\nu\tau,\sigma} + g_{\tau\sigma,\nu} - g_{\nu\sigma,\tau}).
\eeq
The time component of the equation of motion yields
\beq
  \dot{\epsilon} + \frac{\dot{\mu}}{\mu} \epsilon 
   + 2 \frac{\dot{a}}{a} \epsilon \vect{x}^2 = 0
\eeq
with $\epsilon \equiv
\sqrt{\vect{x}^{\prime\,2}/(1-\dot{\vect{x}}^2)}$. On the other hand,
the spatial components yield
\beq
  \ddot{\vect{x}} 
    + 2 \frac{\dot{a}}{a} (1-\dot{\vect{x}}^2) \dot{\vect{x}}
    - \frac{1}{\epsilon} (\epsilon^{-1} \vect{x}')^{\prime} = 0.
\eeq
We define the energy $E$ of a cosmic string in an expanding universe as
\beq
  E = a(\tau) \mu(\tau) \int d\sigma \epsilon.
\eeq
The evolution of the energy density as $\rho \equiv E/V$ with $V$ some
relevant volume is given by
\beq
  \dot{\rho} = -2 \frac{\dot{a}}{a} (1 + \la v^2 \ra) \rho,
\eeq
where $\la v^2 \ra$ is the average velocity squared of a cosmic string
defined as
\bea
  \la v^2 \ra \equiv \la \dot{\vect{x}}^2 \ra  
              \equiv \frac{\int d\sigma \epsilon \dot{\vect{x}}^2}
                          {\int d\sigma \epsilon}.
\eea
Note that multiplying the rate equation of $\rho$ by $d\tau / dt$ gives
an equation with the same form but a derivative taken with respect to
the cosmic time $t$.

The evolution of the string network is characterized by a
correlation length $L$ as
\beq
  \rho_{\infty} = \frac{\mu(t)}{L^2},
  \label{eq:rhoinfty}
\eeq
where $\rho_{\infty}$ is the energy density of a string whose length is
larger than the horizon scale (called infinite strings). In fact,
strings intercommute and their energy is transferred from infinite
strings to loops. The rate of energy transfer from infinite strings to
loops is given by
\beq \dot{\rho}_{\infty \rightarrow {\rm loops}} = \tilde{c} v
  \frac{\rho_{\infty}}{L}, \eeq
where $\tilde{c}$ parametrizes the efficiency of energy transfer and $v
\equiv \sqrt{\la v^2 \ra}$ is the average velocity. Then, the rate
equation for the energy density of infinite strings becomes
\beq
  \frac{d\rho_{\infty}}{dt} = 
    - 2 H (1 + v^2) \rho_{\infty}
    - \tilde{c}v \frac{\rho_{\infty}}{L},
\eeq
where $H$ is the Hubble parameter. Inserting Eq. (\ref{eq:rhoinfty})
into this equation yields
\beq
  \frac{dL}{dt} =
    HL (1 + v^2) + \frac12 \tilde{c} v
    + \frac{L}{2\mu} \frac{d\mu}{dt}.
    \label{eq:Leq}
\eeq
On the other hand, the evolution equation of the velocity is given by
\beq
  \frac{dv}{dt} = (1-v^2) \lmk \frac{\tilde{k}}{L}-2Hv \rmk.
  \label{eq:velocityeq}
\eeq
Here we have taken $\la \dot{\vect{x}}^2 \ra^2 = \la \dot{\vect{x}}^4
\ra$, which is exact up to second order terms. In the one-scale model,
the typical curvature radius is given by the correlation length $L$,
\beq
  \frac{a}{L} \hat{\vect{u}} = \frac{d^2 \vect{x}}{ds^2},
\eeq
where $\hat{\vect{u}}$ is a unit vector and $s$ is the physical length
along the string. As introduced in Ref. \cite{MS}, we have defined the
dimensionless parameter $\tilde{k}$, which is associated with the
presence of small-scale structure,
\beq
  \tilde{k} \equiv  \frac{ \la (1-\dot{\vect{x}}^2) (\dot{\vect{x}} 
                    \cdot \hat{\vect{u}}) \ra }
                    {v(1-v^2)}.
\eeq
Note that the velocity evolution equation (\ref{eq:velocityeq}) is the
same as that for a string with constant tension because the terms
related to time dependence of the tension are cancelled.

In order to investigate the time development of $L$, we define $\gamma$
as $L = \gamma t$ and assume that $\mu \propto t^{q}$. Then,
Eqs. (\ref{eq:Leq}) and (\ref{eq:velocityeq}) can be recast into
\bea
  \frac{1}{\gamma} \frac{d\gamma}{dt}    
     &=& - \frac{1}{2t}
       \lmk 1 - q - v^2 - \frac{\tilde{c} v}{\gamma} \rmk,  \\
  \frac{dv}{dt} &=& 
     \frac{1-v^2}{t} \lmk \frac{\tilde{k}}{\gamma}-v \rmk,
\eea
where we have assumed the radiation domination. These equations have the
stable fixed point $\gamma_{\rm r}$ and $v_{\rm r}$, which is given by
\bea
  \gamma_{\rm r} &=& \frac{\tilde{c}_{\rm r} v_{\rm r}}{1 - q - v^2_{\rm r}}
                  =  \sqrt{
          \frac{\tilde{k}_{\rm r}(\tilde{k}_{\rm r}+\tilde{c}_{\rm r})} 
                     {1-q}}, \\
  v_{\rm r} &=& \frac{\tilde{k}_{\rm r}}{\gamma_{\rm r}}
             = \sqrt{\frac{\tilde{k}_{\rm r}(1-q)}
                          {\tilde{k}_{\rm r}+\tilde{c}_{\rm r}} }.
\eea
In the same way, the stable fixed point in a matter-dominated universe
$\gamma_{\rm m}$ and $v_{\rm m}$ is given by
\bea
  \gamma_{\rm m} &=& \frac{3\tilde{c}_{\rm m} v_{\rm m}}
                     {2 - 3q - 4v^2_{\rm m}}
                  = \frac{3}{2}\sqrt{
          \frac{\tilde{k}_{\rm m}(\tilde{k}_{\rm m}+\tilde{c}_{\rm m})} 
                    {2-3q}}, \\
  v_{\rm m} &=& \frac{3\tilde{k}_{\rm m}}{4\gamma_{\rm m}}
             = \frac12 \sqrt{\frac{\tilde{k}_{\rm m}(2-3q)}
                          {\tilde{k}_{\rm m}+\tilde{c}_{\rm m}} }.
\eea 

Note that the stable fixed point exists only when $q<1$ in the radiation
domination and $q<2/3$ in the matter domination. In case that the stable
fixed point exists, the characteristic scale $L$ of a string network
scales with time for strings with time-dependent tension. That is, the
number of infinite strings per horizon volume is a constant irrespective
of time. But it should be noted that, due to the time dependence of the
tension, the ratio of the energy density of infinite strings to that of
the background universe is {\it not} a constant.

It is convenient to work with the comoving correlation length $l \equiv
L/a$ and the conformal time $\tau$. Then, the evolution equations are
rewritten as
\bea
  \dot{l} &=& \frac{\dot{a}}{a} l v^2 + \frac12 \tilde{c}v 
            + \frac{l}{2} \frac{\dot{\mu}}{\mu}, \\
  \dot{v} &=& (1-v^2) \lmk \frac{\tilde{k}}{l}
                           - 2 \frac{\dot{a}}{a} v \rmk.
\eea
According to Ref. \cite{PV}, we interpolate between these values through the
radiation-matter transition,
\bea
\tilde{c}(\tau)=\frac{\tilde{c}_r+g a \tilde{c}_m}{1+g a}, \\
\tilde{k}(\tau)=\frac{\tilde{k}_r+g a \tilde{k}_m}{1+g a},
\eea
where we take $\tilde{c}_r=0.23$, $\tilde{c}_m=0.18$,
$\tilde{k}_r=0.17$, $\tilde{k}_m=0.49$, $g=300$ and $a(\tau)$ is
normalized so that $a=1$ at present \cite{PV}. Here, we expect that the
time dependence of the tension has little effect on the parameters
$\tilde{c}$ and $\tilde{k}$ because intercommutation is a local process
near the string core and $\tilde{k}$ is a parameter associated with the
presence of small-scale structure. For simplicity, we set the wiggleness
parameter $\alpha$ to be unity.

\section{Constraints on the string tension}

In this section, we first show CMB and matter power spectra seeded by
cosmic strings whose tension varies with time in addition to those by
the constant tension strings which are conventionally
investigated. Then we discuss cosmological constraints on the
tension. As is well known for the constant tension case, since they
show no acoustic oscillation and positive correlation in
temperature-polarization cross-correlation spectrum (TE), initial
perturbation from cosmic strings alone cannot explain the observed CMB
spectra and the inflationary adiabatic initial perturbation
necessarily dominates.  However, some contribution from cosmic string
is still allowed.  Therefore, we derive upper bounds on the string
tension by studying how much the string contribution can be compared
with the adiabatic one. We use the WMAP three-year data \cite{WMAP}
for CMB and SDSS data release 2 \cite{SDSS} for galaxy clustering data
to give the constraints. We denote the cosmological parameters as follows:
 baryon density $\omega_b$, matter 
density $\omega_m$, hubble parameter $h$, reionization optical depth $\tau$,
spectral index of primordial spectrum $n_s$.

\begin{figure}[htb]
\vspace{1cm}
\begin{center}
\scalebox{0.6}{
\includegraphics{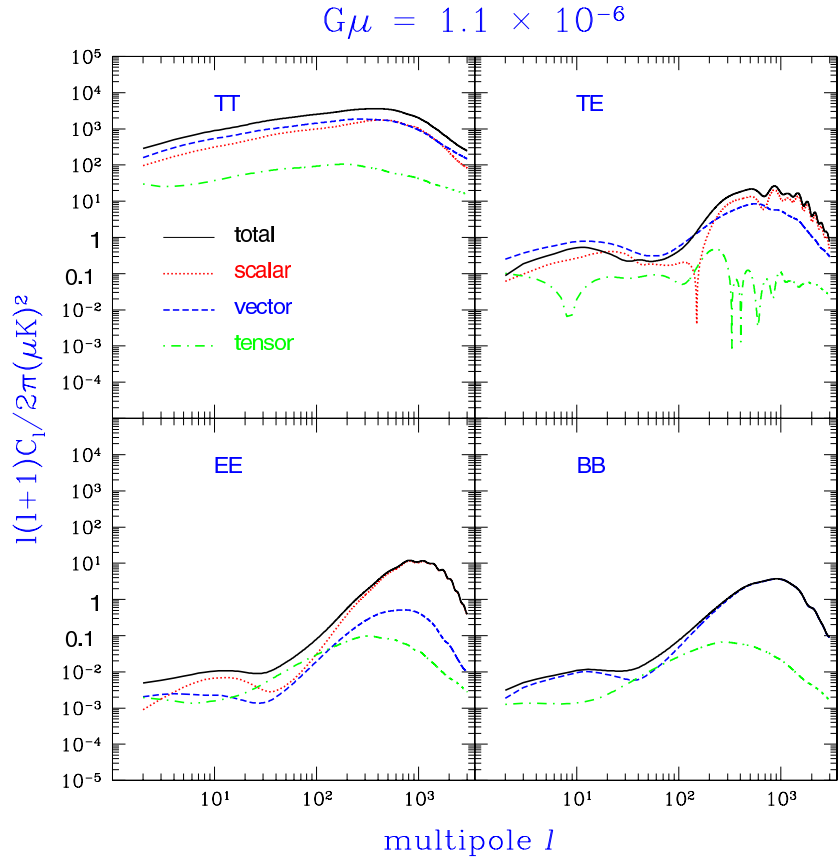}}
\vspace{1cm}
\caption{CMB power spectrum for the case with constant tension.}
\label{fig:cl_pow0}
\end{center}
\end{figure}

\begin{figure}[htb]
\begin{center}
\scalebox{0.6}{
\includegraphics{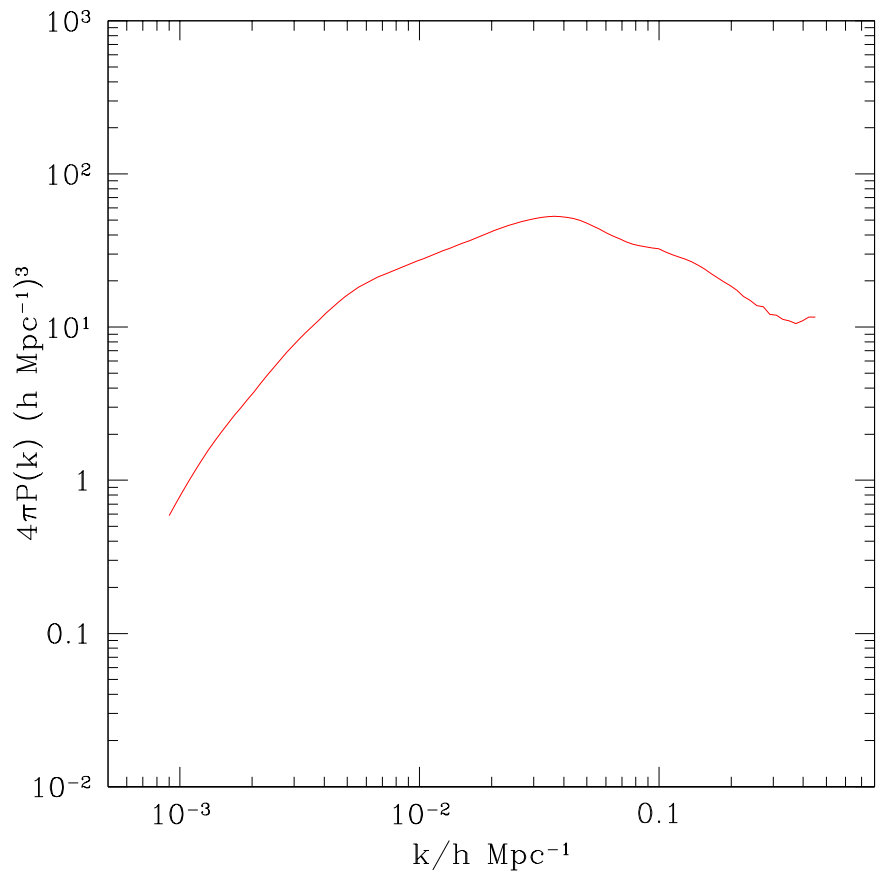}}
\caption{Matter power spectrum for the case with constant tension.}
\label{fig:Pk_pow0}
\end{center}
\end{figure}

\begin{figure}[t]
\begin{center}
\scalebox{0.38}{\includegraphics{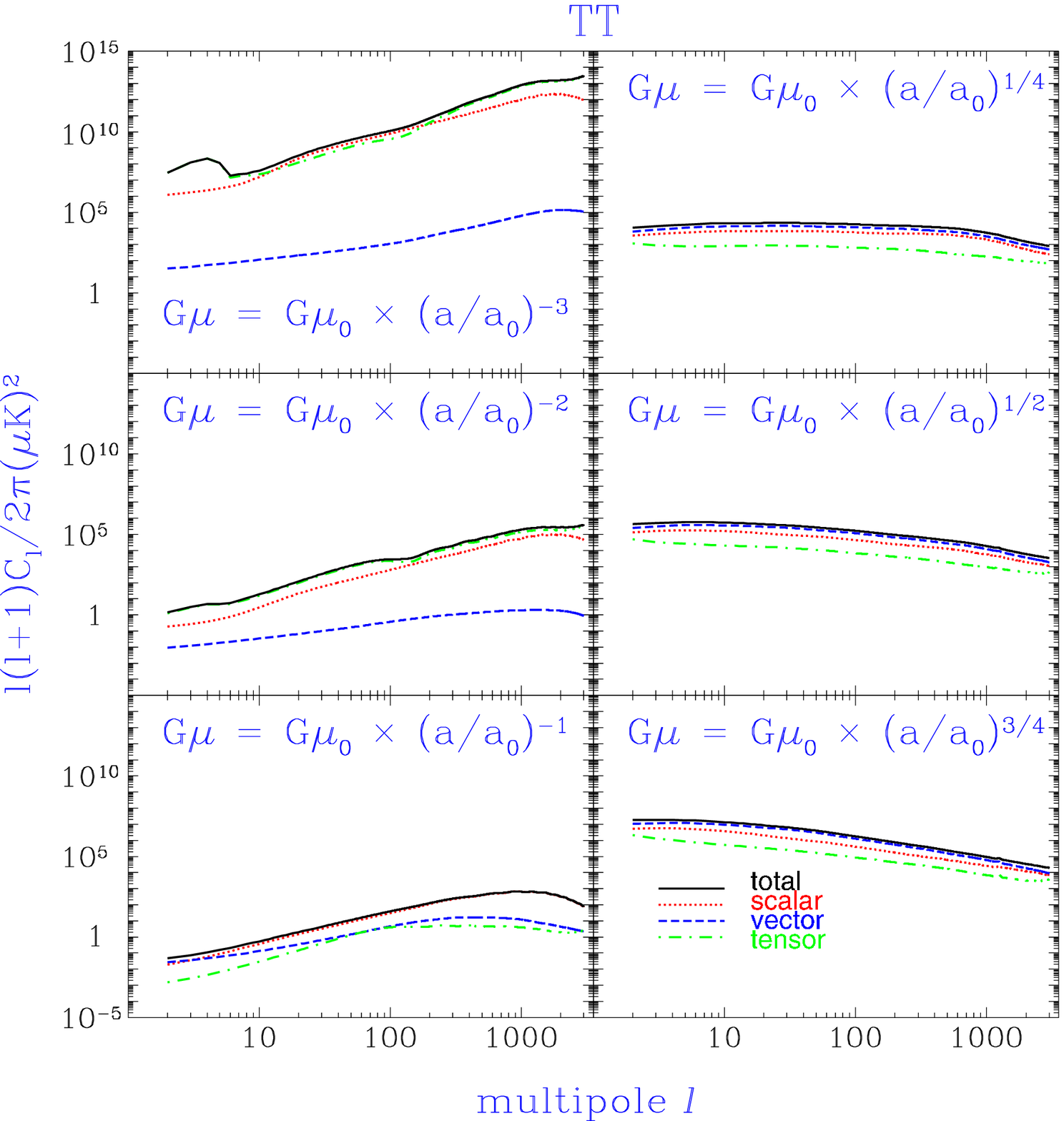}} 
\hspace{1cm}
\scalebox{0.38}{\includegraphics{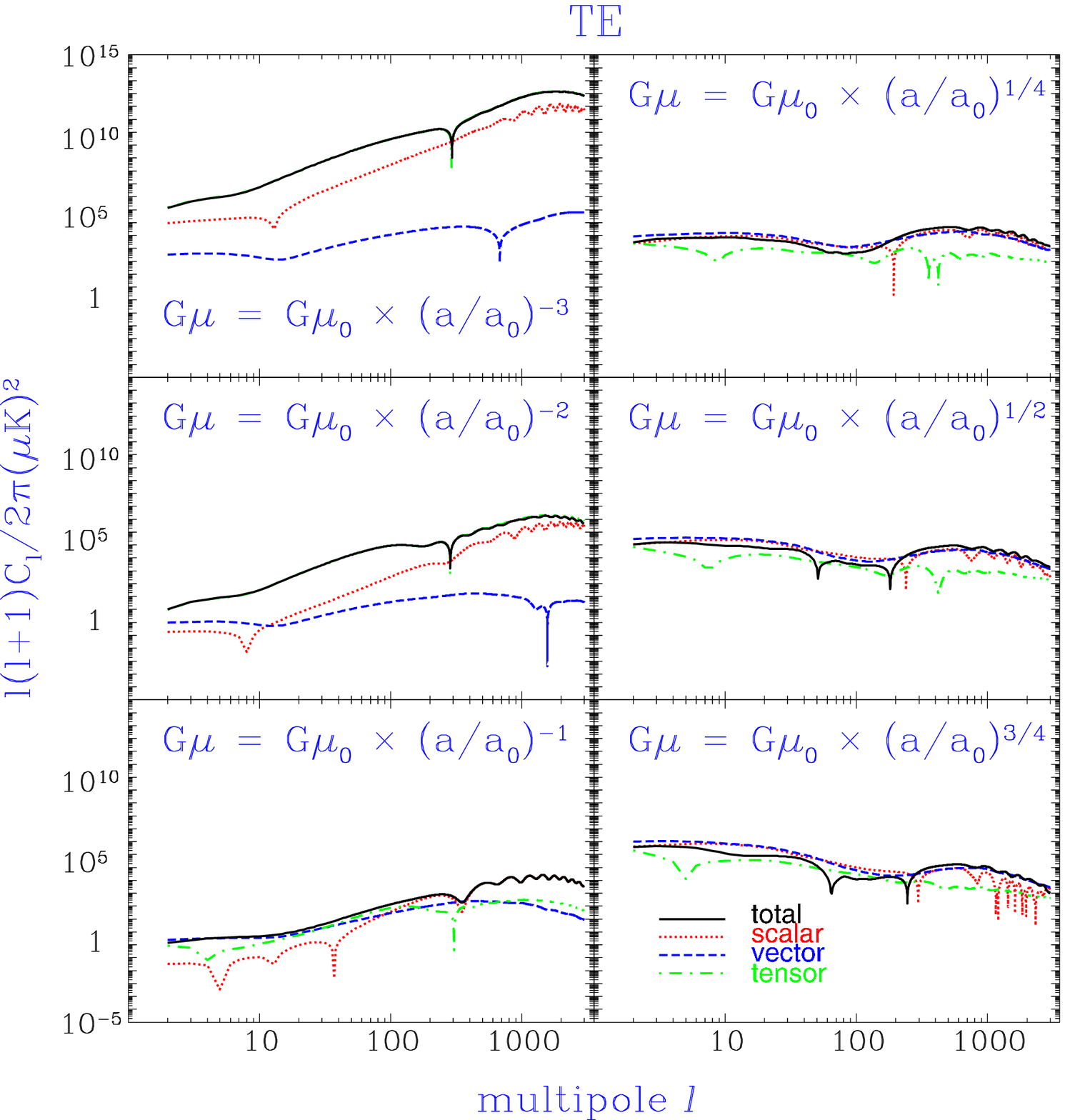}} \\
\vspace{1cm}
\scalebox{0.38}{\includegraphics{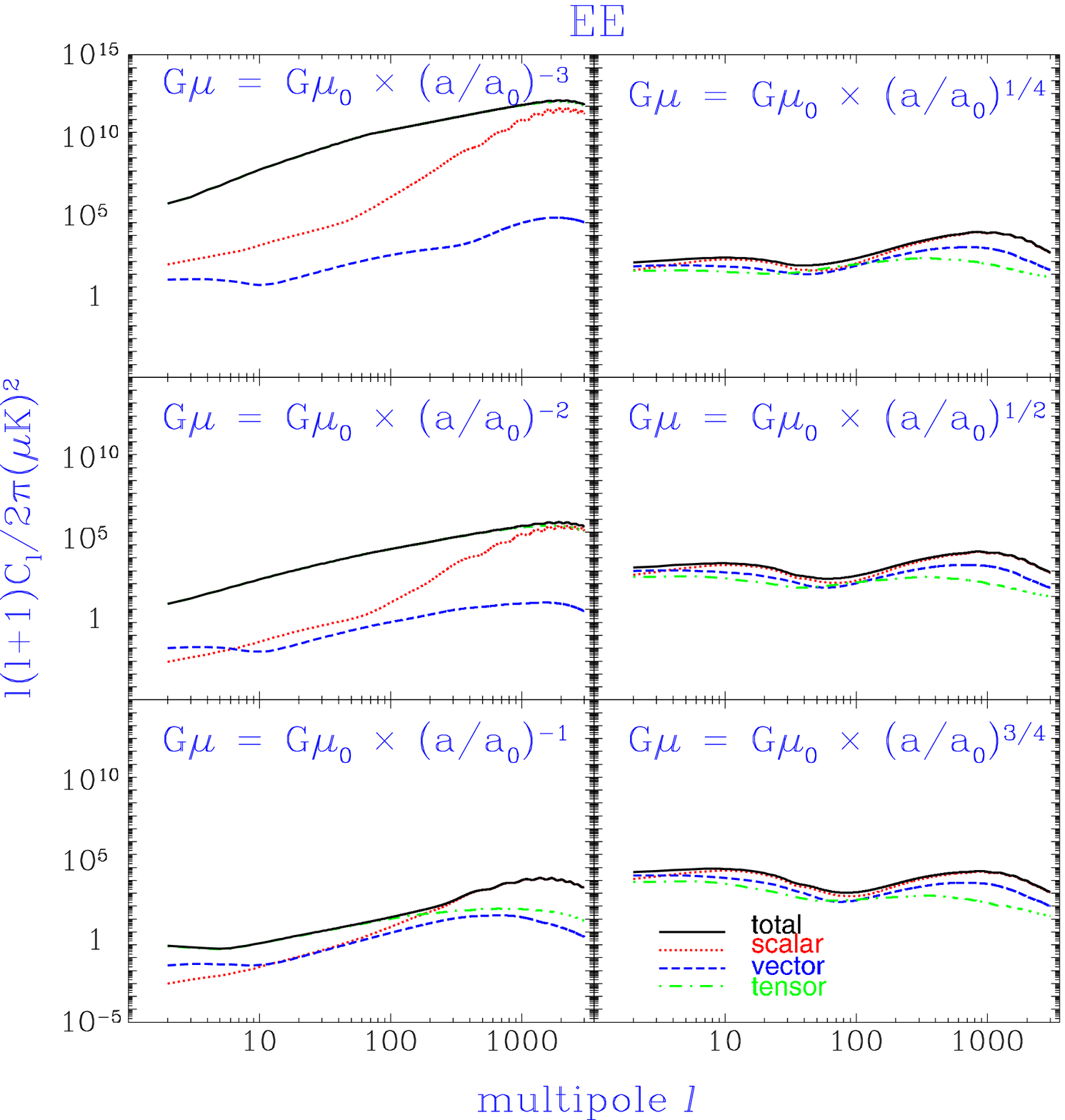}} 
\hspace{1cm}
\scalebox{0.38}{\includegraphics{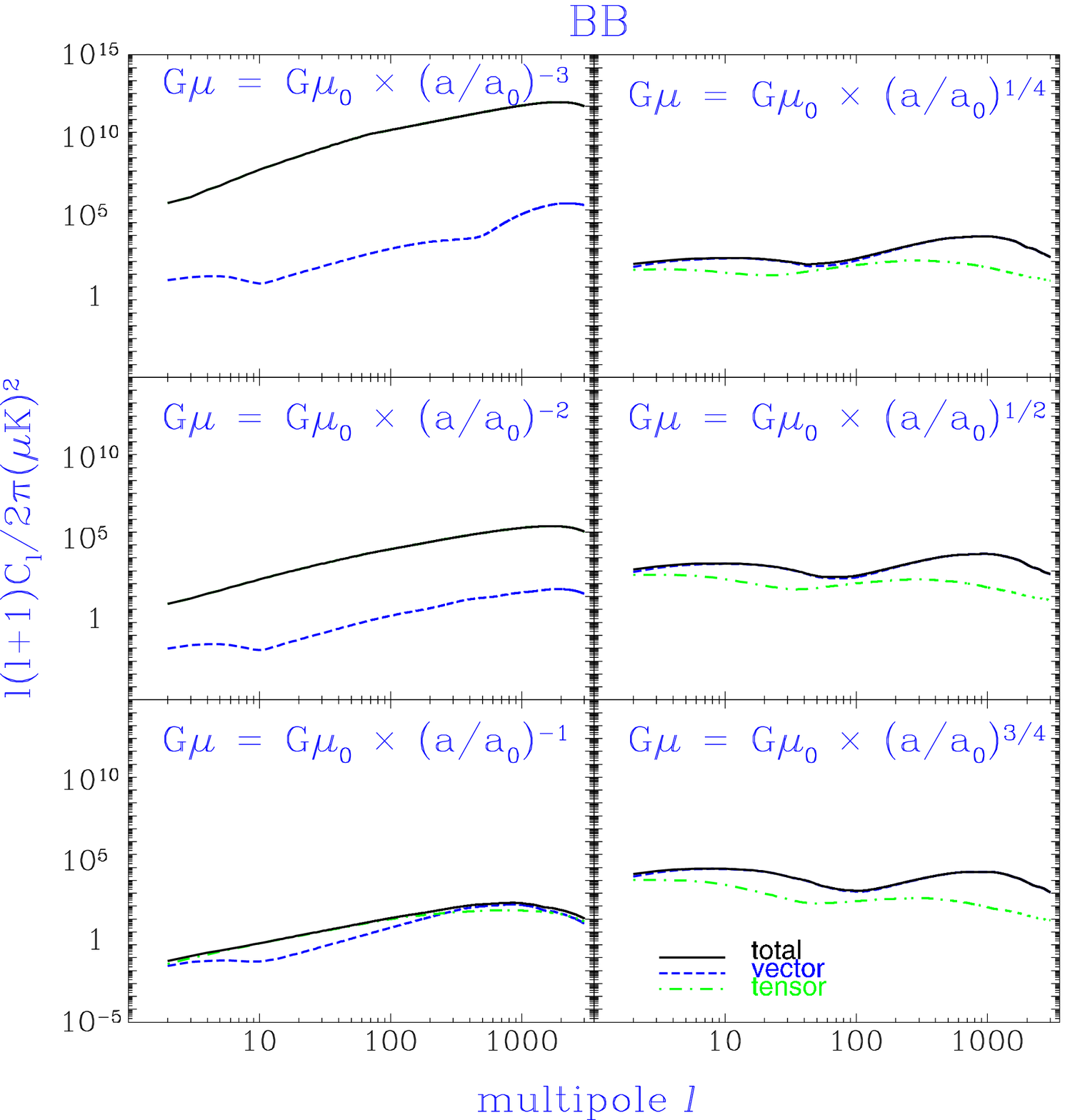}}
\vspace{1cm}
\caption{CMB TT, TE, EE, BB power spectra 
  for the cases with time varying tension $G\mu \propto a^n$, $n=-3, -2,
  -1, 1/4, 1/2, 3/4$. Scalar, vector, tensor modes and the total power spectrum are shown. Note that the absolute
  magnitude is given for TE power spectra.}
\label{fig:aCMB}
\end{center}
\end{figure}

\begin{figure}[t]
\begin{center}
\scalebox{0.5}{\includegraphics{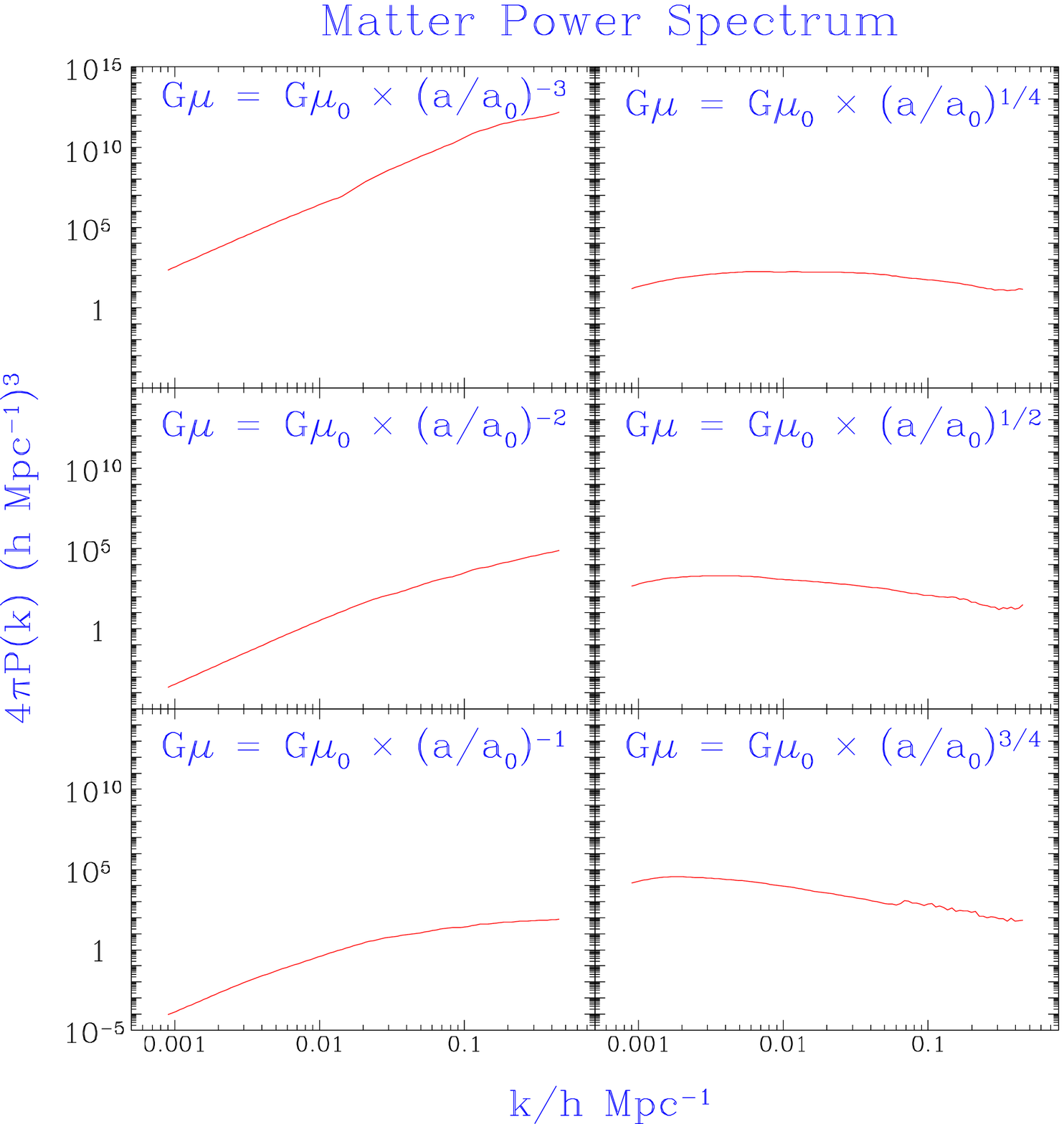}}
\vspace{1cm}
\caption{Matter power spectra for the cases with time varying tension
  $G\mu \propto a^n$, $n=-3, -2, -1,  1/4, 1/2, 3/4$.}
\label{fig:aMP}
\end{center}
\end{figure}

Let us begin with showing the results of power spectra calculation of
cosmic strings whose tension $\mu$ varies in proportion to some power
$n$ of the scale factor $a$, $\mu \propto a^n$.  For comparison, first
we show the CMB and matter power spectra for the case with cosmic
strings with constant tension in Figs.~\ref{fig:cl_pow0} and
\ref{fig:Pk_pow0} respectively.  Figure~\ref{fig:aCMB} and
\ref{fig:aMP} are for the cases with time dependent tension with
negative powers ($n=-3, -2, -1$) and with positive powers ($n=1/4,
1/2, 3/4$). Note that, as shown above, the string network does not
follow a scaling law for $n \gtrsim 1$. These spectra are calculated
by modifying CMBACT code \cite{PV,PWW} which is based on CMBFAST code
\cite{Seljak:1996is}. In these figures, the cosmological parameters
are taken to be the WMAP mean values $\omega_b = 0.0223$, $\omega_m =
0.127$, $h=0.73$, $\tau = 0.09$ \cite{WMAP}.

\begin{figure}[t]
\begin{center}
\scalebox{0.9}{\includegraphics{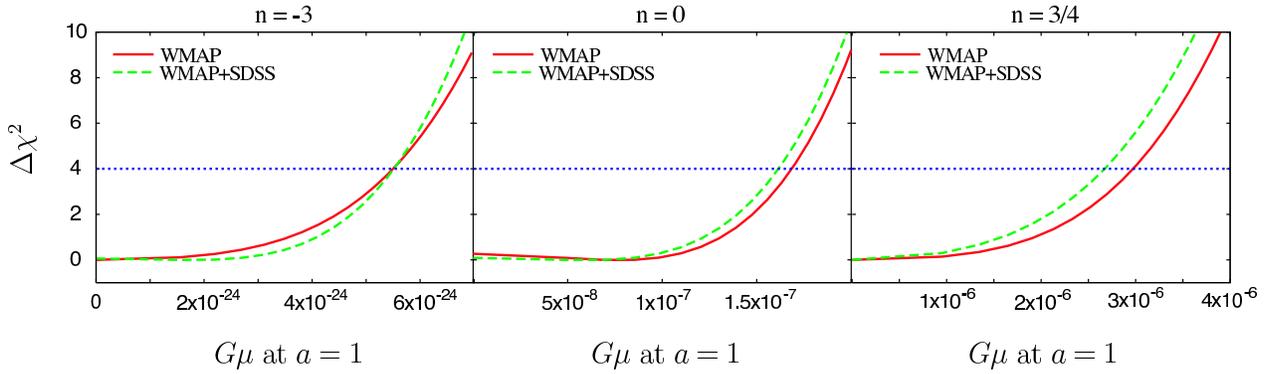}}
\caption{The values of $\Delta \chi^2$ from WMAP only and WMAP+SDSS
  are shown as a function of $G\mu$ at the present time. Here we
  assume the time dependence of $G\mu$ as $G\mu \propto a^n$. The
  cases with $n=-3, 0$ and $3/4$ are shown.}
\label{fig:chi2}
\end{center}
\end{figure}

We use these spectra sourced by cosmic strings to place upper bound on
the string tension. After adding power spectra from adiabatic initial
perturbation, we plug the sum into the likelihood codes for CMB and
galaxy clustering supplied respectively by the WMAP group
\cite{Jarosik:2006ib,Hinshaw:2006ia,Page:2006hz} and the SDSS group
\cite{SDSS}. For each value of the string tension, we calculate
$\chi^2$ by marginalizing over the amplitude of adiabatic power
spectrum and the bias factor between observed galaxy power spectrum
and theoretical matter power spectrum (the sum of adiabatic and string
contribution). In connection with the latter, we use the galaxy
clustering measurements in 19 $k$-bands with $k < 0.2 h/$Mpc and
assume that the bias factor does not depend on $k$. Other cosmological
parameters are fixed to be $\omega_b = 0.0221$, $\omega_m =
0.127$, $h = 0.725$, $\tau= 0.0913$, $n_s = 0.957$ for
deriving WMAP alone constraints and $\omega_b = 0.0225$, $\omega_m =
0.145$, $h = 0.664$, $\tau= 0.0816$ and $n_s = 0.959$ for
WMAP plus SDSS constraints, which minimize $\chi^2$ of respective data
sets when the cosmic string contribution is absent ($i.e.$ when only
adiabatic initial perturbation exists). This minimization is carried
out by the method introduced in Ref.~\cite{Ichikawa:2004zi} and the
minimum $\chi^2$ for WMAP alone case agrees with the WMAP analysis as
noted in Ref.~\cite{Fukugita:2006rm}. We show the values of $\Delta
\chi^2$ as a function of $G\mu$ at the present time for some cases in
Fig.~\ref{fig:chi2}.  In table \ref{tab:constraint_a}, we report upper
bounds at 95\% confidence level by reading $G\mu$ which gives $\Delta
\chi^2 =4$. While the bounds on string tensions at present become
stringent for negative powers, they are weakened for positive
powers. The inclusion of the SDSS data does not improve the
constraints significantly.

Note that our omission of full marginalization over the cosmological
parameters might underestimate upper bounds, but we note that we have
obtained the constraint from WMAP and SDSS for the constant tension case
to be $G\mu < 1.6 \times 10^{-7}$ in good agreement with the bound $G\mu
< 2.7 \times 10^{-7}$ obtained by Pogosian, Wasserman and Wyman
\cite{PWW}\footnote{This constraint is obtained from the WMAP first year
results, while ours are from the WMAP three year results.}. Our result
for the case with constant tension also agrees with those obtained in Refs.~\cite{Fraisse} and \cite{Seljak}.  This shows that our rather simplified
way of estimating the bounds works well.

\begin{figure}[t]
\begin{center}
\scalebox{0.38}{\includegraphics{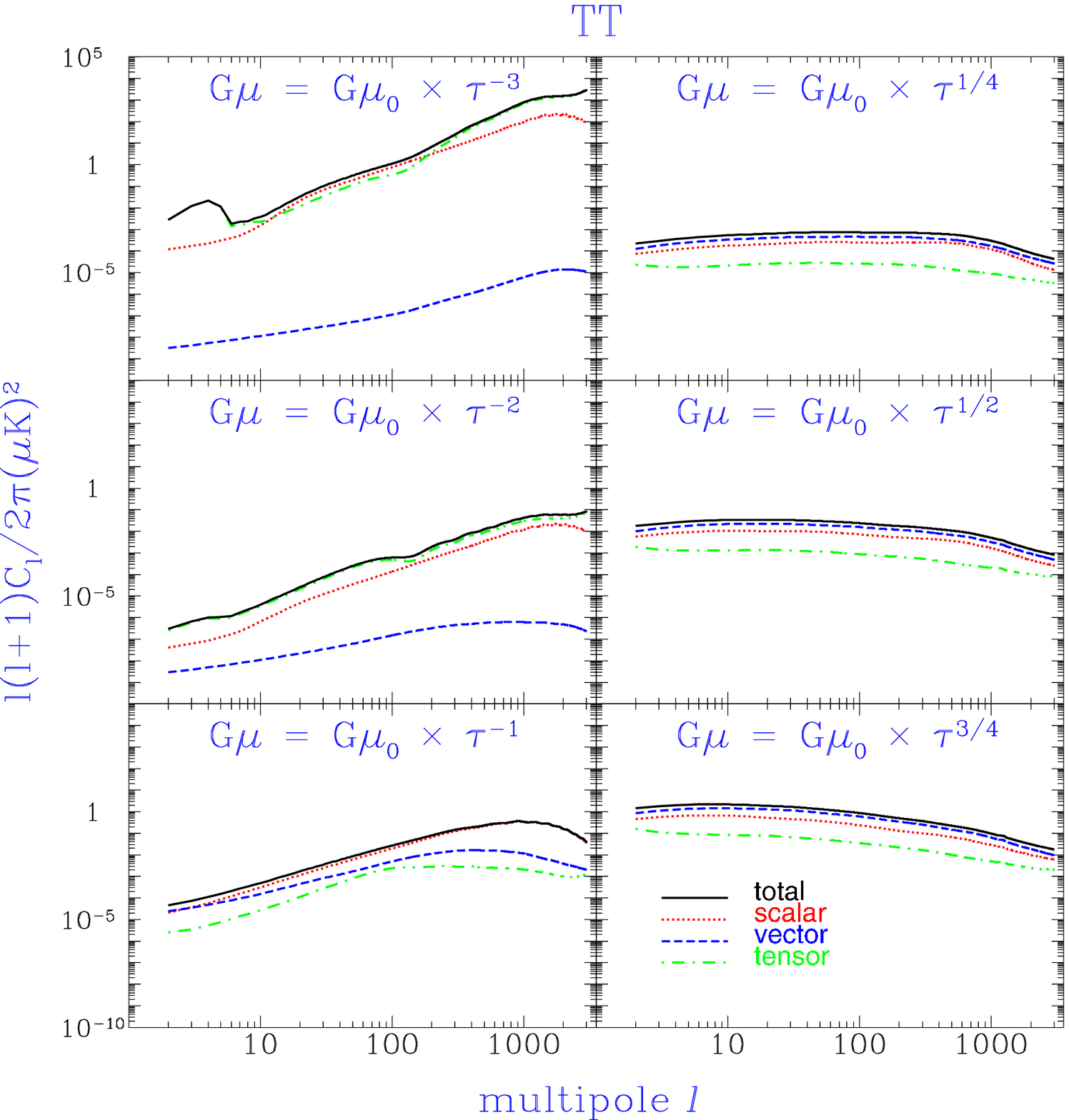}} 
\hspace{1cm}
\scalebox{0.38}{\includegraphics{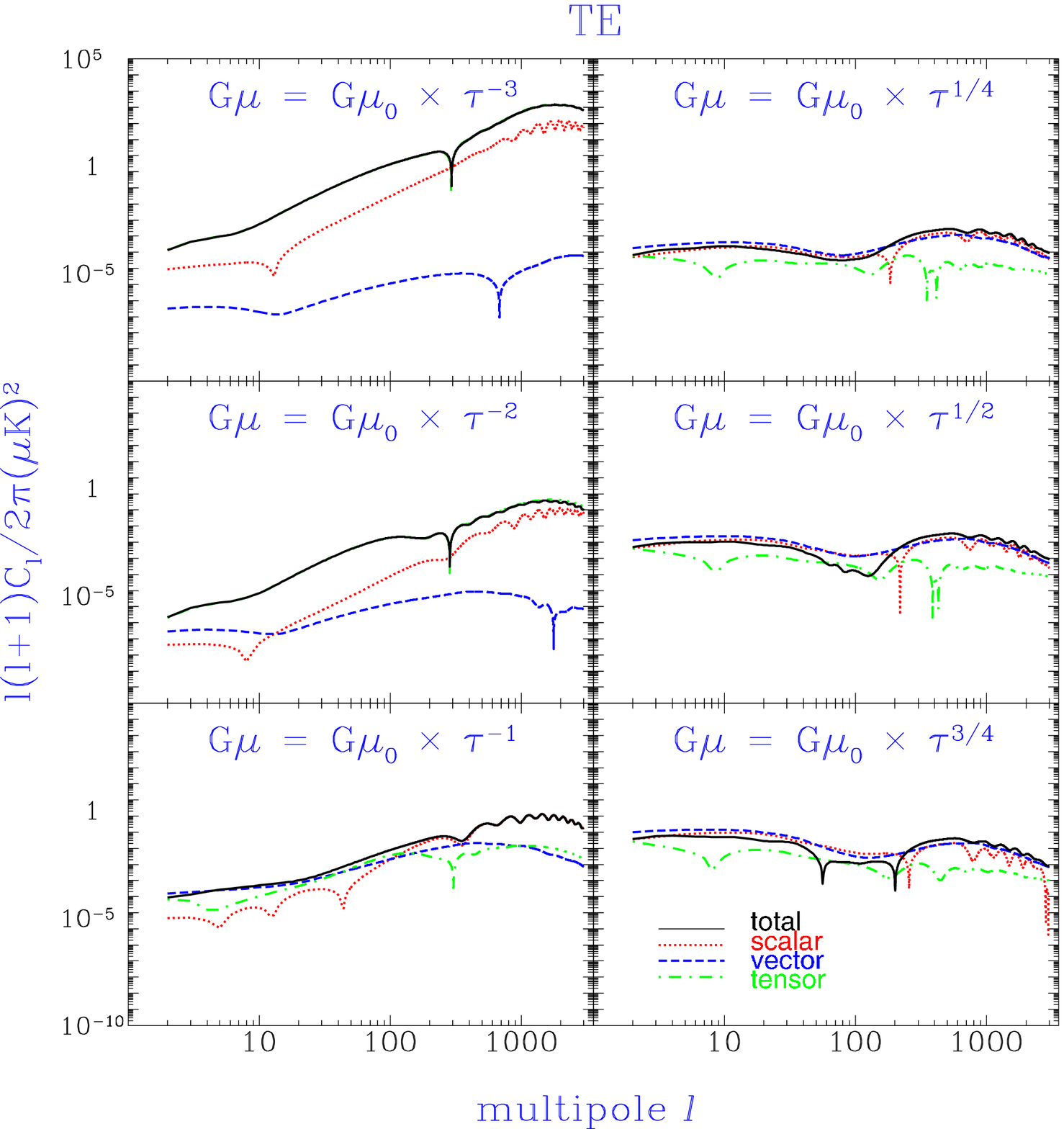}} \\
\vspace{1cm}
\scalebox{0.38}{\includegraphics{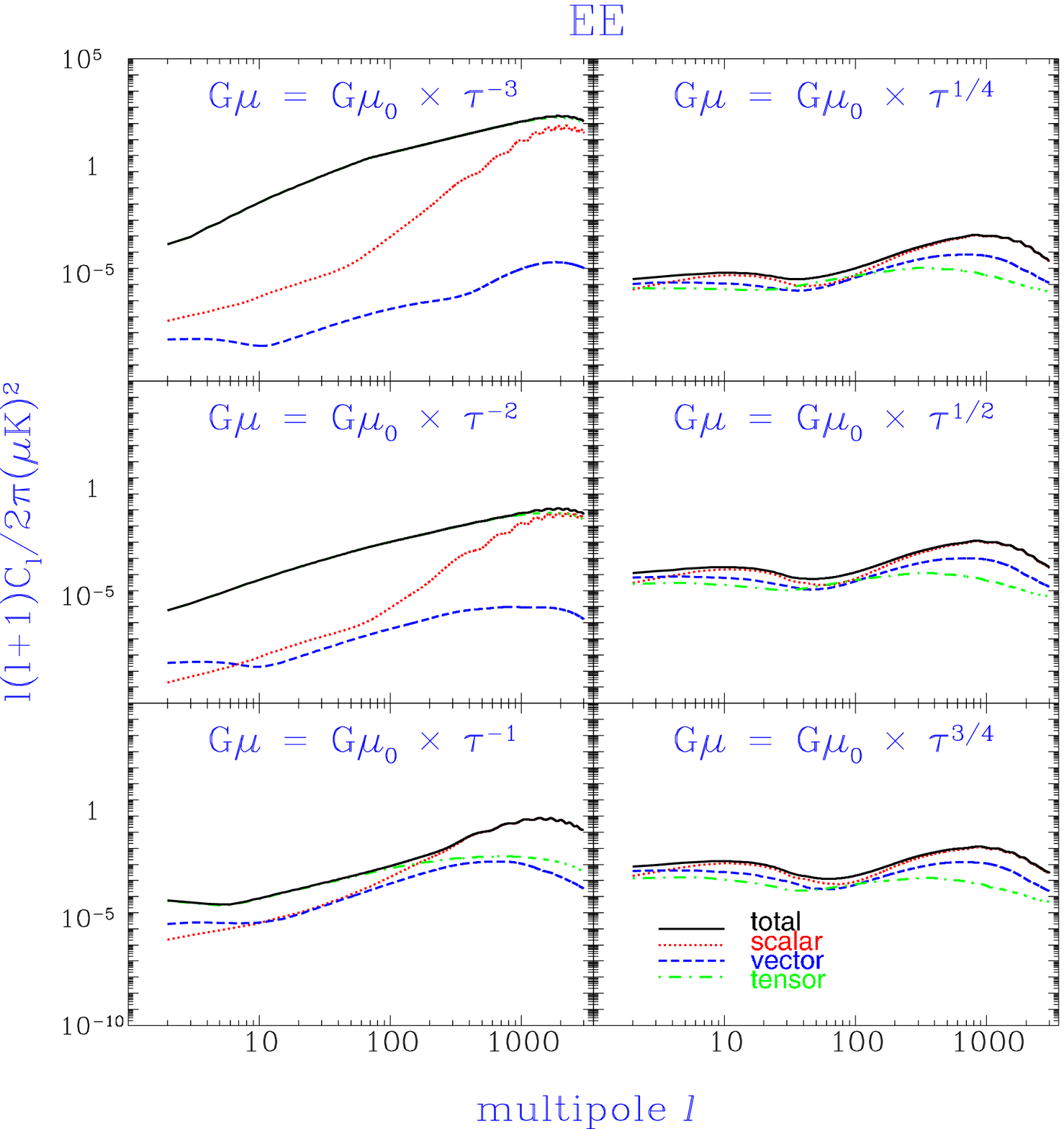}} 
\hspace{1cm}
\scalebox{0.38}{\includegraphics{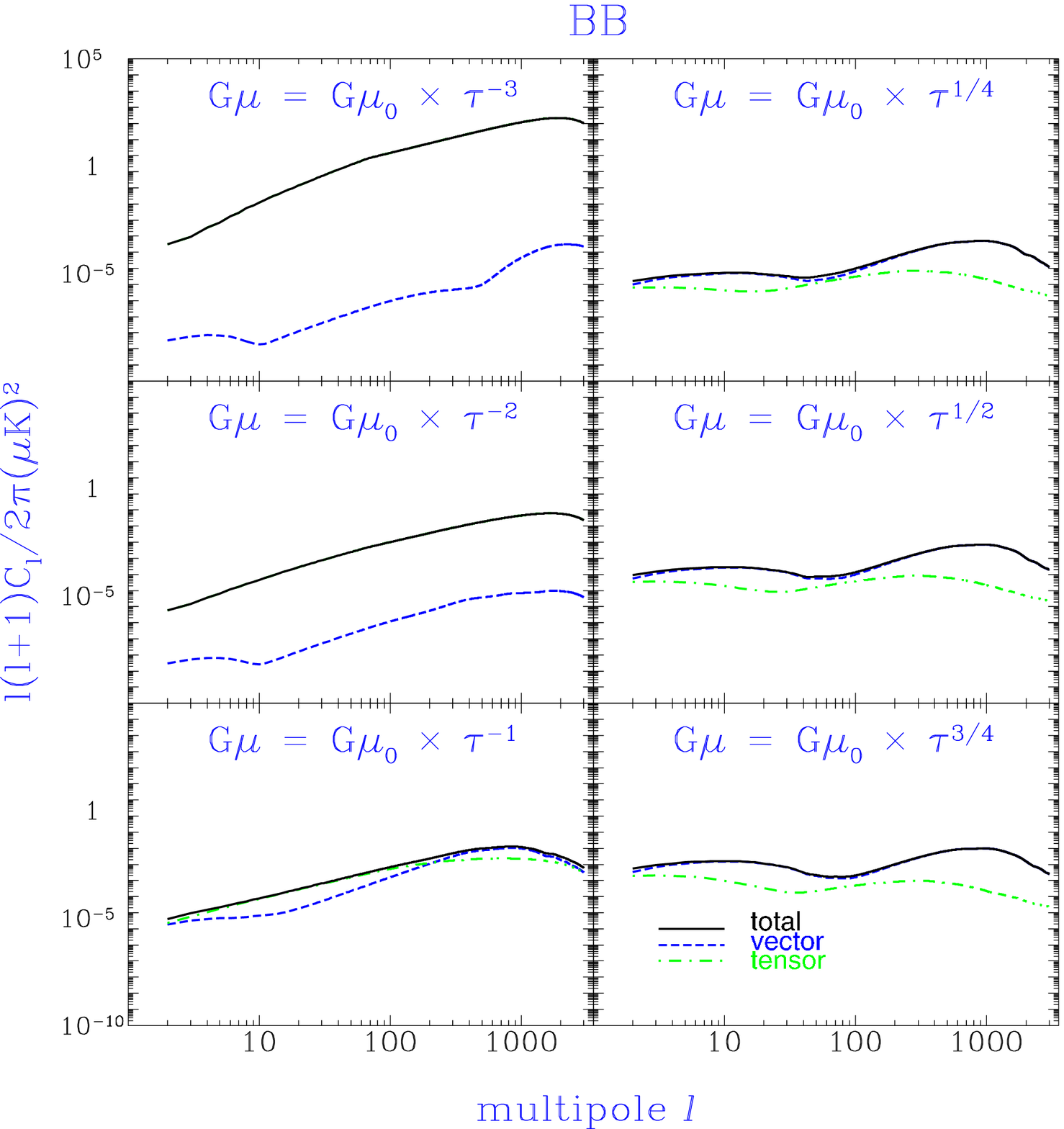}}
\vspace{1cm}
\caption{CMB TT, TE, EE, BB power spectra 
  for the cases with time varying tension $G\mu \propto \tau^n$, $n=-3,
  -2, -1, 1/4, 1/2, 3/4$.  Scalar, vector, tensor modes and the total power spectrum are shown. Note that the absolute magnitude is given for TE power
  spectra.}
 \label{fig:tauCMB}
\end{center}
\end{figure}

\begin{figure}[t]
\begin{center}
\scalebox{0.5}{\includegraphics{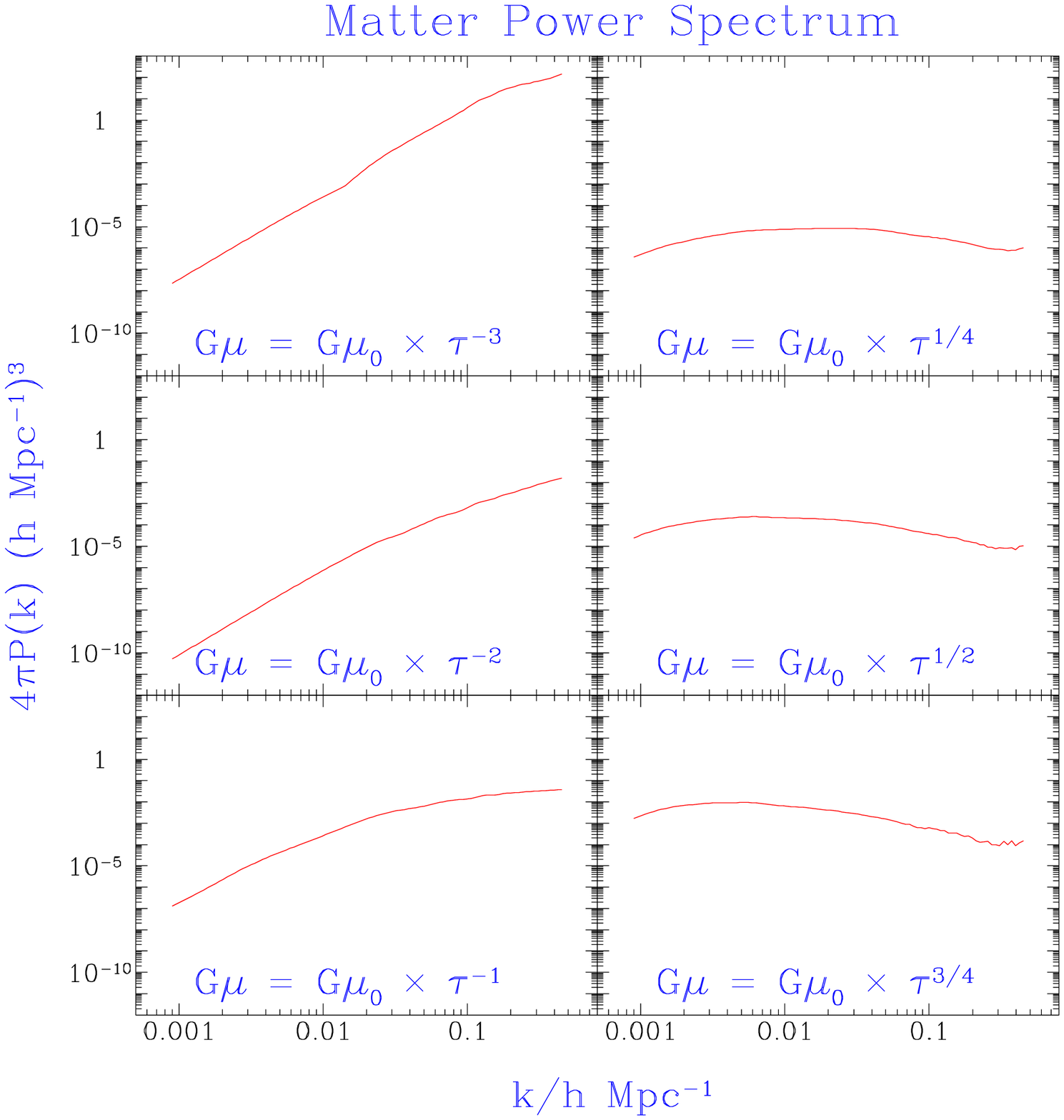}}
\vspace{1cm}
\caption{Matter power spectra for the cases with time varying tension
  $G\mu \propto \tau^n$, $n=-3, -2, -1,  1/4, 1/2, 3/4$.}
\label{fig:tauMP}
\end{center}
\end{figure}

We repeat similar analysis with the cosmic strings whose tension scales
as a power law of the conformal time $\tau$, $\mu \propto \tau^n$. The
power spectra are shown in figures~\ref{fig:tauCMB} and \ref{fig:tauMP},
and constraints are summarized in table~\ref{tab:constraint_tau}. We
have obtained constraints which show similar trend to the case of a
power law of $a$. They become significantly severer for the cases of
negative powers.

\section{Conclusions and Discussion}

In this paper, we have discussed cosmological evolution and implications
of cosmic strings with time-dependent tension. First of all, we have
derived equations of motions based on the velocity-dependent one scale
model in the expanding universe. By using these equations, we
investigate whether cosmic strings with time-dependent tension go into
the scaling solution when the tension depends on some power of the
cosmic time $\mu \propto t^q$. Then, we find that such strings relax
into the scaling solution only when $q<1$ in the radiation domination
and $q<2/3$ in the matter domination.

Then we showed the CMB and matter power spectra sourced by cosmic
strings with time-dependent tension. As shown in the previous section,
the spectra can be different significantly from those produced by the
conventional cosmic strings with constant tension.  We have also
discussed the constraints on the time-dependent tension from the
Wilkinson microwave anisotropy probe (WMAP) three year data and Sloan
digital sky survey (SDSS) data.  Since the time dependence of the
tension strongly depends on models, here we considered it in some
general settings.  For the time dependence of string tensions, we
assumed that the tension $\mu$ changes as some power of the scale factor
$a$, $\mu \propto a^n$, and the conformal time $\tau$, $\mu \propto
\tau^n$.  The constraints on $\mu$ for various values of $n$ were given.
For the cases with negative powers, the bounds on string tensions at
present time become much stronger than those for the case with constant
tension.  One should, however, notice that the tension can be large in
the past in this case, which may have implications on the structure
formation at very small scales. On the other hand, the bounds are
weakened for positive powers, which may lead to large amount of
gravitational radiation background. We will explorer these possibilities
in future work.

\section*{Acknowledgments}

We would like to acknowledge the use of CMBACT code developed and made
publicly available by L. Pogosian and T. Vachaspati. M.Y. is supported
in part by the project of the Research Institute of Aoyama Gakuin
University and by the JSPS Grant-in-Aid for Scientific Research No.\
18740157.

\pagebreak 

\centerline{Tables:}

\vspace{1cm}

\begin{table}[htb]
\begin{tabular}{|c|c|c|}
\hline 
$n$ & WMAP alone (95\%) & WMAP+SDSS (95\%) \\
\hline
-3 & $5.5 \times 10^{-24}$& $5.5 \times 10^{-24}$\\
-2 & $2.8 \times 10^{-16}$& $2.9 \times 10^{-16}$\\
-1 & $3.7 \times 10^{-11}$ & $3.8 \times 10^{-11}$\\
0 & $1.7 \times 10^{-7}$ &$1.6 \times 10^{-7}$\\
1/4 & $6.4 \times 10^{-7}$& $5.8 \times 10^{-7}$\\
1/2 & $1.4 \times 10^{-6}$& $1.2 \times 10^{-6}$\\
3/4 & $3.0 \times 10^{-6}$& $2.7 \times 10^{-6}$\\
\hline
\end{tabular}
\caption{Constraints for constant and time varying ($G\mu \propto a^n$,
  $n=-3, -2, -1, 0, 1/4, 1/2, 3/4$) tension. We quote the upper bounds
  on $G\mu$ at the present epoch ($a=1$).}
\label{tab:constraint_a}
\end{table}

\vspace{3cm}

\begin{table}[htb]
\begin{tabular}{|c|c|c|}
\hline 
$n$ & WMAP alone (95\%) & WMAP+SDSS (95\%) \\
\hline
-3 & $1.8 \times 10^{-19}$& $1.8 \times 10^{-19}$\\
-2 & $2.9 \times 10^{-13}$& $3.0 \times 10^{-13}$\\
-1 & $1.1 \times 10^{-9}$ & $1.1 \times 10^{-9}$\\
0 & $1.7 \times 10^{-7}$ &$1.6 \times 10^{-7}$\\
1/4 & $4.0 \times 10^{-7}$& $3.8 \times 10^{-7}$\\
1/2 & $6.3 \times 10^{-7}$& $5.6 \times 10^{-7}$\\
3/4 & $9.0 \times 10^{-7}$& $8.2 \times 10^{-7}$\\
\hline
\end{tabular}
\caption{Constraints for constant and time varying ($G\mu \propto
\tau^n$, $n=-3, -2, -1, 0, 1/4, 1/2, 3/4$) tension. We quote the upper
bounds on $G\mu$ at the present epoch ($\tau =1.5 \times 10^4$).}
\label{tab:constraint_tau}
\end{table}

\pagebreak 

\centerline{Figures:}

\begin{thebibliography}{9}


\bib{Kibble}
T. W. B. Kibble, 
\JP{9}{1387}{76}.

\bib{KVY}
L. A. Kofman and A. D. Linde, 
\NPB{282}{555}{87};
Q. Shafi and A. Vilenkin, 
\PRD{29}{1870}{84};
E. T. Vishniac, K. A. Olive, and D. Seckel, 
\NPB{289}{717}{87};
J. Yokoyama, 
\PLB{212}{273}{88}; 
\PRL{63}{712}{89}.

\bib{WMAP}
C.~L.~Bennett {\it et al.},
\APJSS{148}{1}{03};
D. N. Spergel {\it et al.},
astro-ph/0603449.

\bib{PWW}
L. Pogosian, S. -H. H. Tye, I. Wasserman and M. Wyman,
\PRDD{68}{023506}{03};
L. Pogosian, M. Wyman, and I. Wasserman,
\JCAPP{09}{008}{04};
M. Wyman, L. Pogosian, and I. Wasserman,
\PRDD{72}{023513}{05}; 
\IBB{73}{089905}{06};
L. Pogosian, M. Wyman, and I. Wasserman,
astro-ph/0604141.

\bib{Fraisse}
A. A. Fraisse,
astro-ph/0503402.

\bib{Seljak}
U.~Seljak, A.~Slosar and P.~McDonald,
arXiv:astro-ph/0604335.

\bib{VSHK}
For review, see A. Vilenkin and E. P. S. Shellard, 
{\it Cosmic String and Other Topological Defects}
(Cambridge University Press, Cambridge, England, 1994);
M. B. Hindmarsh and T. W. B. Kibble,
\RPP{58}{477}{95}.

\bib{ER}
P. P. Avelino and A. R. Liddle,
\MNRASS{348}{105}{04};
P. P. Avelino and D. Barbosa,
\PRDD{70}{067302}{04};
L. Pogosian and A. Vilenkin,
\PRDD{70}{063523}{04};
K. D. Olum and A. Vilenkin;
astro-ph/0605465. 

\bib{GR}
A. Vilenkin,
\PLBold{107}{47}{81};
T. Damour and A. Vilenkin,
\PRDD{71}{063510}{05}.

\bib{GL}
M. Sazhin {\it et al.},
\MNRASS{343}{353}{03};
R. Schild {\it et al.},
\AAA{422}{477}{04}.

\bib{BMY}
R. H. Brandenberger, A. Mazumdar, and M. Yamaguchi,
\PRDD{69}{081301(R)}{04}.

\bib{CSS}
E. Witten,
\PLBold{153}{243}{85}.

\bib{BI}
N. Jones, H. Stoica, and S. -H. H. Tye,
\JHEPP{07}{051}{02};
S. Sarangi and S. -H. H. Tye,
\PLBB{536}{185}{02};
N. T. Jones, H. Stoica, S. -H. H. Tye,
\PLBB{563}{6}{03}.

\bib{MY}
M. Yamaguchi,
\PRDD{72}{043533}{05}.

\bib{AT}
A. Albrecht and N. Turok,
\PRL{54}{1868}{85};
A. Albrecht and N. Turok,
\PRD{40}{973}{89}.

\bib{BB}
D. P. Bennett and F. R. Bouchet,
\PRL{60}{257}{88};
\IB{63}{2776}{89};
\PRD{41}{2408}{90}.

\bib{AS}
B. Allen and E. P. S. Shellard,
\PRL{64}{119}{90}.

\bib{NG} 
Y. Nambu, in {\it Proceedings of International Conference on
  Symmetries \& Quark Models} Lectures at the Copenhagen Summary
Symposium, 1970; T. Goto, \PTP{46}{1560}{71}.

\bib{gs}
M. Yamaguchi, J. Yokoyama, M. Kawasaki,
\PTP{100}{535}{98};
M. Yamaguchi, M. Kawasaki, and J. Yokoyama,
\PRL{82}{4578}{99};
M. Yamaguchi, 
\PRD{60}{103511}{99};
M. Yamaguchi, J. Yokoyama, and M. Kawasaki,
\IBIDD{61}{061301(R)}{00};
M. Yamaguchi and J. Yokoyama,
\IBIDD{66}{121303(R)}{02};\IBB{67}{103514}{03}.

\bib{gm}
M. Barriola and A. Vilenkin,
\PRL{63}{341}{89};
D. P. Bennett and S. H. Rhie,
\IBID{65}{1709}{90};
U. Pen, D. N. Spergel, and N. Turok,
\PRD{49}{692}{94};
M. Yamaguchi,
\PRDD{64}{081301(R)}{01}; \IBB{65}{063518}{02}.

\bib{MS}
C. J. A. P. Martins and E. P. S. Shellard,
\PRD{53}{575(R)}{96};
\IB{54}{2535}{96}.

\bib{SDSS}
M Tegmark {\it et al.},
\APJJ{606}{702}{04}.

\bib{PV}
L. Pogosian and T. Vachaspati,
\PRD{60}{083504}{99}.

\bibitem{Seljak:1996is}
U.~Seljak and M.~Zaldarriaga,
Astrophys.\ J.\  {\bf 469}, 437 (1996)

\bibitem{Jarosik:2006ib}
    N.~Jarosik {\it et al.},
    arXiv:astro-ph/0603452.

\bibitem{Hinshaw:2006ia}
    G.~Hinshaw {\it et al.},
    arXiv:astro-ph/0603451.

\bibitem{Page:2006hz}
    L.~Page {\it et al.},
    arXiv:astro-ph/0603450.

\bibitem{Ichikawa:2004zi}
  K.~Ichikawa, M.~Fukugita and M.~Kawasaki,
  Phys.\ Rev.\ D {\bf 71}, 043001 (2005).

\bibitem{Fukugita:2006rm}
  M.~Fukugita, K.~Ichikawa, M.~Kawasaki and O.~Lahav,
  arXiv:astro-ph/0605362.

\end{thebibliography}
\end{document}